\documentclass[a4paper,11pt]{article}
\usepackage{jinstpub} 
\usepackage{lineno}

\title{\boldmath Co-Design of 2D Heterojunctions for Data Filtering in Tracking Systems}

\author[1]{Tupendra Oli,\note{Corresponding author.}}
\author[1]{Wilkie Olin-Ammentorp,}
\author[1]{Xingfu Wu,}
\author[2]{Justin H. Qian,}
\author[2]{Vinod K. Sangwan,}
\author[2]{Mark C. Hersam,}
\author[1]{Salman Habib,}
\author[1]{and Valerie Taylor}

\affiliation[1]{Argonne National Laboratory, Lemont, IL 60439, USA}
\affiliation[2]{Northwestern University, Evanston, IL 60208, USA}

\emailAdd{toli@anl.gov}

\abstract
{As particle physics experiments evolve to achieve higher energies and resolutions, handling the massive data volumes produced by silicon pixel detectors, which are used for charged particle tracking, poses a significant challenge. To address the challenge of data transport from high-resolution tracking systems, we investigate a support vector machine (SVM)-based data classification system designed to reject low-momentum particles in real-time. This SVM system achieves high accuracy through the use of a customized `mixed' kernel function, which is specifically adapted to the data recorded by a silicon tracker. Moreover, this custom kernel can be implemented using highly efficient, novel van der Waals heterojunction devices. This study demonstrates the co-design of circuits with applications that may be adapted to meet future device and processing needs in high-energy physics (HEP) collider experiments.
}


\begin{document}
\maketitle

\section{Introduction}
\label{sec:intro}
Experiments such as ATLAS~\cite{alam2008atlas} and CMS~\cite{collaboration2008cms} at the Large Hadron Collider (LHC) are at the forefront of scientific discovery in high-energy particle physics. These experiments collect proton-proton (\textit{pp}) collision data at the LHC to study fundamental particles, their interactions, and to search for new particles and phenomena at the energy frontier. The discovery of the Higgs boson in 2012 marked a significant milestone in the history of physics \cite{DellaNegra_Jenni_Virdee_2012}. These experiments collect enormous amounts of data, reaching tens of terabytes per second from collisions occurring every 25 nanoseconds \cite{Das_2022}. To push the limits of discovery potential, the LHC is being upgraded to the High-Luminosity LHC (HL-LHC), while the HEP community is also envisioning and designing even more powerful next-generation colliders, such as the Future Circular Collider (FCCee, FCChh)~\cite{fcc2019fcc, fcc2019hh}, to address critical unanswered questions. These future experiments will produce unprecedented data rates in the order of petabytes per second, making efficient data processing and management a significant challenge for the next generation of experiments.

\begin{figure}[htbp]
 \centering
 \includegraphics[width= 0.8 \linewidth]{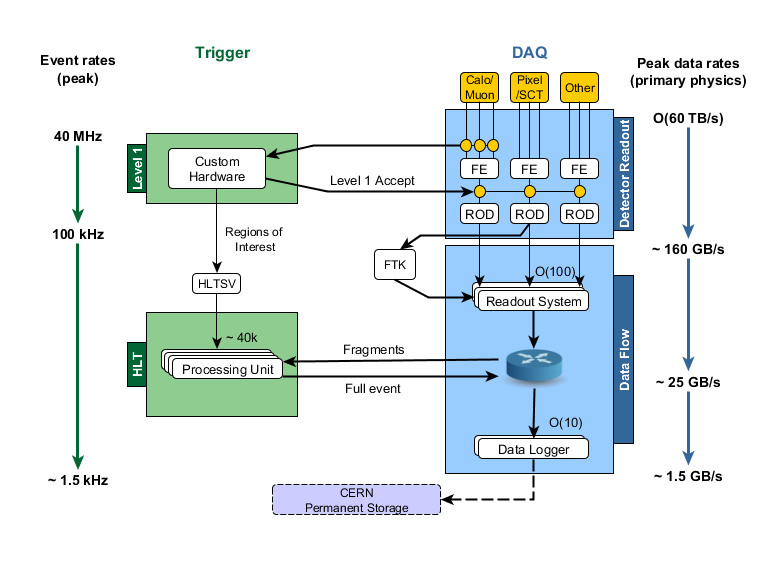}
 \caption{Overview of the ATLAS Trigger Data Acquisition (TDAQ) architecture, showing data rates and bandwidths at each level of the trigger system, adapted from \cite{tdaq}.}
 \label{fig:fig0}
 \end{figure}

Particle detectors are highly segmented, comprising multiple layers of detection systems—from the innermost tracking detector to the electromagnetic and hadronic calorimeters, and finally the muon detectors. The total data production rate generated by all detector systems is immense - the current data rate for ATLAS and CMS across the entire system is approximately 40 terabytes per second, which is too large to read directly from the detectors and store for offline analysis \cite{Das_2022}. However, only a small portion of this data is relevant for further analysis in the search for new physics and particles. To determine which events to keep and which to discard, two stages of processing are employed as shown in Figure~\ref{fig:fig0}: first, a hardware-based trigger system implemented on custom electronics boards reduces the data rate to around 100 kHz, and then the software-based High-Level Trigger (HLT) further reduces it to about 1 kHz. However, with the increased luminosity and energy in future colliders, the initial data rate will rise significantly, necessitating innovative capabilities in the experiments. 

The tracking detectors, which consist of silicon pixel and strip sensors, are crucial for capturing the primary interactions during beam collisions. They provide essential data for tracking, vertexing, and flavor tagging. With the upgrades in future detectors, the amount of data produced by tracking in particular is increasing~\cite{fcc2019hh}. The information from the tracker is useful for identifying particles based on their trajectories. However, the sheer volume of data makes integration into the trigger system challenging. As a result, tracking is not included in the hardware-based trigger system and is currently restricted to the software-based HLT system or done in offline reconstruction in both ATLAS~\cite{vazquez2016atlas} and CMS \cite{Das_2022}. 

In the pixel detector within the tracker, a significant portion of the hits is caused by particles with low transverse momentum (low-$p_T$). These low-$p_T$ tracks generally carry less useful information and can be filtered out before the data is transmitted off the detector. Various on-pixel data reduction strategies have been conceptualized. These include smart pixel implementations on ASICs~\cite{yoo2024smart}, and FPGA-based tracklets~\cite{bartz2020fpga}. However, each approach presents challenges, primarily because they rely on digital hardware, which is facing well-known challenges in continued miniaturization. Additionally, these systems must be engineered to withstand the high radiation levels in collider experiments. 

In this paper, we investigate the feasibility of a novel SVM-based two-dimensional (2D) van der Waals heterojunction device that operates in the analog domain and offers unique advantages, including high energy efficiency and a low radiation cross-section. Section~\ref{sec:svm} discusses the SVM algorithm and mixed kernel approach in detail. Section~\ref{sec:mkp} discusses the use of mixed-kernel approach in charge cluster classification in tracker of HEP experiments and its performance. In Section~\ref{sec:svmd}, we explore the energy consumption and potential co-design opportunities within the trigger system. Finally, Section~\ref{sec:discuss} summarizes our findings and outlines directions for future work.

\section{Support Vector Machines}
\label{sec:svm}
Support Vector Machines (SVMs) are supervised machine learning algorithms, offering robust capabilities in classification, regression, and outlier detection.~\cite{boser1992training,cortes1995support,awad2015support,gunn1998support}. SVMs have demonstrated remarkable performance across a diverse range of classification tasks, finding applications in field such as medical diagnosis, finance, and high energy physics~\cite{zhang2008text,yang2004biological,lin2011large,shin2005application,chen2011support,ahmad2018performance, vannerem1999classifying,vaiciulis2003support,aaltonen2012search,sahin2016performance}. SVMs find the optimal hyperplane that distinguishes between different classes of data. In Sections~\ref{subsec:lcp} and~\ref{subsec:nlcp}, we will describe how the SVM algorithm addresses both linear and non-linear classification problems.

\subsection{Linear Classification Problem}
\label{subsec:lcp}
\begin{figure}[htbp]
 \centering
 \includegraphics[width= 0.48 \linewidth]{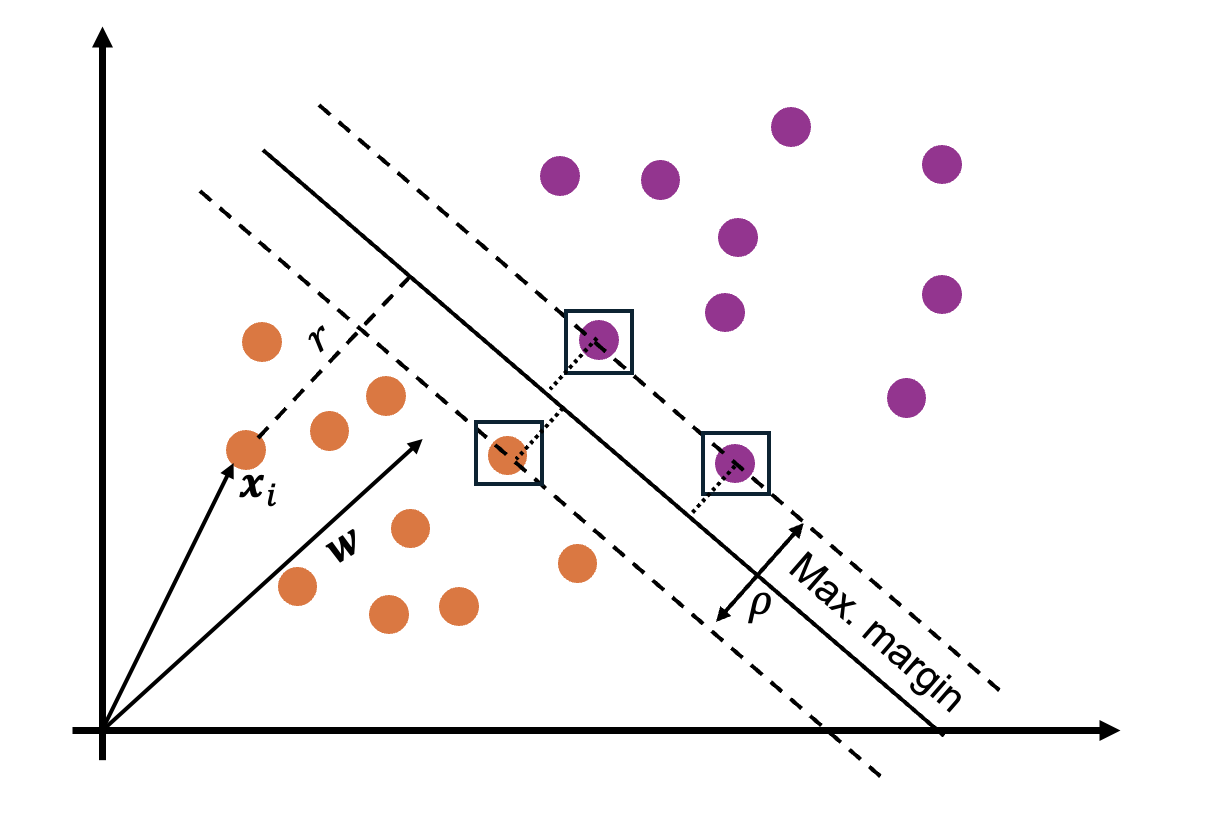}
  \includegraphics[width= 0.48 \linewidth]{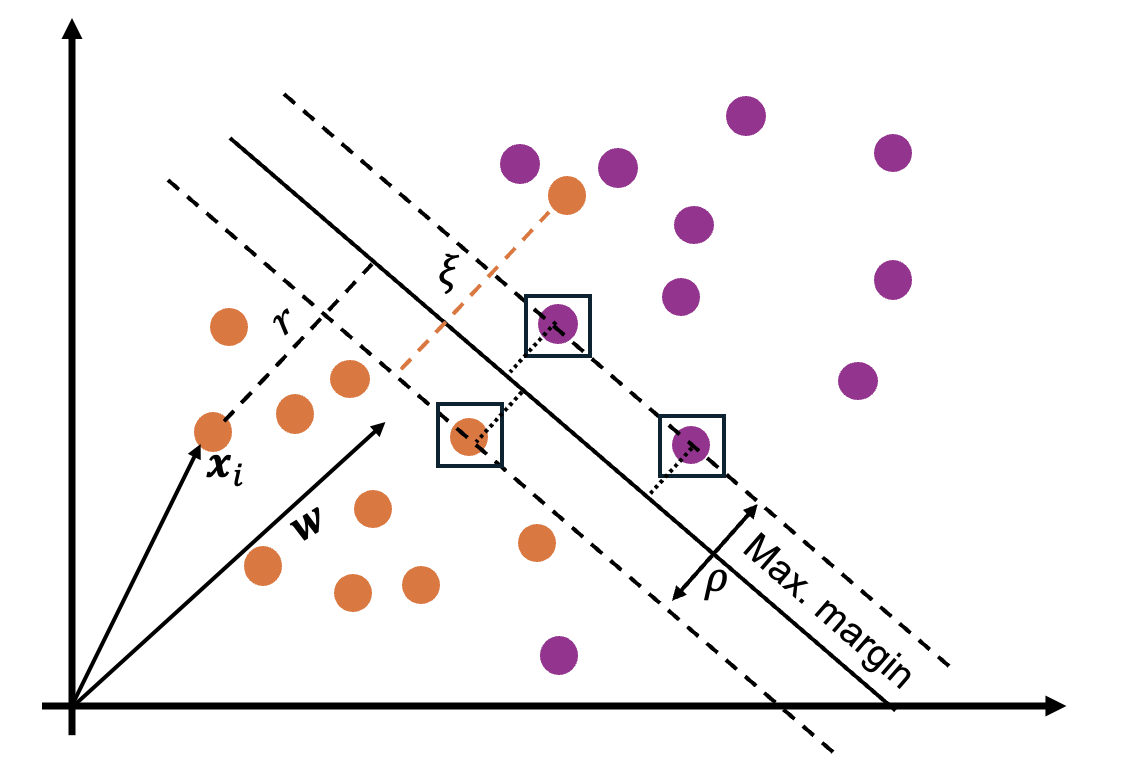}
 \caption{An illustration of SVM classification strategies. On the left, the hard margin SVM classification shows perfect separation of classes with a maximal margin, denoted by $\rho$, between the closest data points (highlighted within squares) and the separating hyperplane. On the right, the soft margin SVM classification allows for some misclassifications, represented by slack variables $\xi_i$, to handle overlapping data and outliers. In both diagrams, $\textbf{x}_{i}$ represents the $i^{th}$ data point vectors, and $\textbf{w}$ is a vector orthogonal to the hyperplane.}
 \label{fig:fig1}
 \end{figure}

For linearly separable data, SVMs employ two main strategies: hard margin and soft margin as represented in Figure~\ref{fig:fig1}. Hard margin SVMs assume a perfect separation between classes and seek to establish a hyperplane with a maximal margin that strictly separates the classes without any overlap. This method is highly sensitive to outliers as it requires all data points to be correctly classified. In contrast, soft margin SVMs introduce a slack variable, allowing for a trade-off between a certain degree of misclassification and high sensitivity to outliers in the underlying data. This flexibility enables the handling of overlapping data and outliers, thereby improving the model's robustness and generalization capabilities \cite{hearst1998support}.

In Figure~\ref{fig:fig1}, the dataset pairs are denoted as $(\textbf{x}_i, y_i)$, where $\textbf{x}_i \in \mathbb{R}^{n}$ represents the feature vector in an n-dimensional space and $y_i \in {(-1, 1)}$ denotes the corresponding class labels, indicating either a negative or positive class. The distance of the vector $\textbf{x}_i$ from the separator (r) is calculated as $\frac{\textbf{w} \cdot \textbf{x}_i + b}{||\textbf{w}||}$, and the maximum margin ($\rho$) is $\frac{2}{||\textbf{w}||}$. The separating condition of the SVM is $y_{i}(\textbf{w}.\textbf{x}_{i} + b) \geq 1$. Here, \textbf{w} and b are referred to as the weight vector and bias terms, respectively, for the given set of training samples, which are to be determined. A fundamental aspect of SVM is identifying the maximum margin hyperplane, which is heavily influenced by specific data points known as 'support vectors.' These support vectors are essential in determining the hyperplane's position, effectively deciding on which side of the hyperplane a new test case will fall. As illustrated in Figure~\ref{fig:fig1}, modifying these support vectors can consequently alter the position of the hyperplane.

In hard margin SVM, the parameters \textbf{w} and b are determined by solving a constrained optimization problem, as outlined in Equation~\ref{eq:eq1}. Maximizing the margin $\rho$ is equivalent to minimizing the function $\phi(\textbf{w})$, while ensuring that the separating inequality conditions are satisfied. This optimization problem is addressed using Lagrange multipliers to find the optimum of $\phi(\textbf{w})$ under the given conditions. The corresponding Lagrangian is formulated in Equation~\ref{eq:eq2}, and the optimal conditions are derived from its differentiation, as described in references \cite{cortes1995support,scholkopf2002learning,shawe2011review}. The final form of the Lagrangian, $\mathcal{L}$, is presented in Equation~\ref{eq:eq3}, where $\alpha_i=0$ for non-support vectors. The decision function for an unknown data sample \textbf{u} is expressed in Equation~\ref{eq:eq4}. However, it is important to note that the computational time required for this optimization scales quadratically with the number of samples ($N^2$). As a result, this approach can become inefficient for large datasets, necessitating the use of approximate optimization methods.

\begin{equation}
\label{eq:eq1}
    \phi(\textbf{w}) = \frac{1}{2}||\textbf{w}||^{2}, \quad y_{i}(\textbf{w}.\textbf{x}_{i} + b) - 1 = 0
\end{equation}

\begin{equation}
    \label{eq:eq2}
    \mathcal{L}(\textbf{w},b,\alpha_{i}) = \frac{1}{2}||\textbf{w}||^{2} - \sum_{i} \alpha_{i} y_{i}[(\textbf{w}.\textbf{x}_{i} + b) - 1], \quad \alpha_{i} \geq 0
\end{equation}

\begin{equation}
    \label{eq:eq3}
    \mathcal{L} = \sum \alpha_{i} -\frac{1}{2}\sum_{i}\sum_{j}\alpha_{i}\alpha_{j}y_{i}y_{j}\textbf{x}_{i}.\textbf{x}_{j}
\end{equation}

\begin{equation}
    \label{eq:eq4}
  f(x) = \sum \alpha_{i}y_{i}\textbf{x}_{i}.\textbf{u} + b 
 \end{equation}

The soft margin SVM classifier is employed in cases of linearly separable datasets with some overlap between the distributions. To allow some misclassification, soft margin SVM utilizes the slack variables as suggested by~\cite{cortes1995support, suykens2001nonlinear}. Consequently the optimization problem is slightly modified to Equation~\ref{eq:eq5}. The solution to the optimization problem is obtained by differentiating the Lagrangian as in hard margin SVM.  The hyperplane is predominantly determined by the support vectors as $\alpha_{i}$ are nonzero exclusively for support vectors data points. Consequently, the decision function, which is used to classify an unknown sample $\textbf{u}$, remains consistent with the function in Equation~\ref{eq:eq4} satisfying the inequality of $0 \leq \alpha_{i} \leq C, \forall \alpha_{i}$

\begin{equation}
\label{eq:eq5}
    \phi(\textbf{w}) = \frac{1}{2}||\textbf{w}||^{2} +C\sum \xi_{i}, \quad y_{i}(\textbf{w}.\textbf{x}_{i} + b) - 1 + \xi_{i} = 0, \quad \xi_{i} \geq 0
\end{equation}

\subsection{Non-linear Classification Problem}
\label{subsec:nlcp}
\begin{figure}[htbp]
 \centering
 \includegraphics[width= \linewidth]{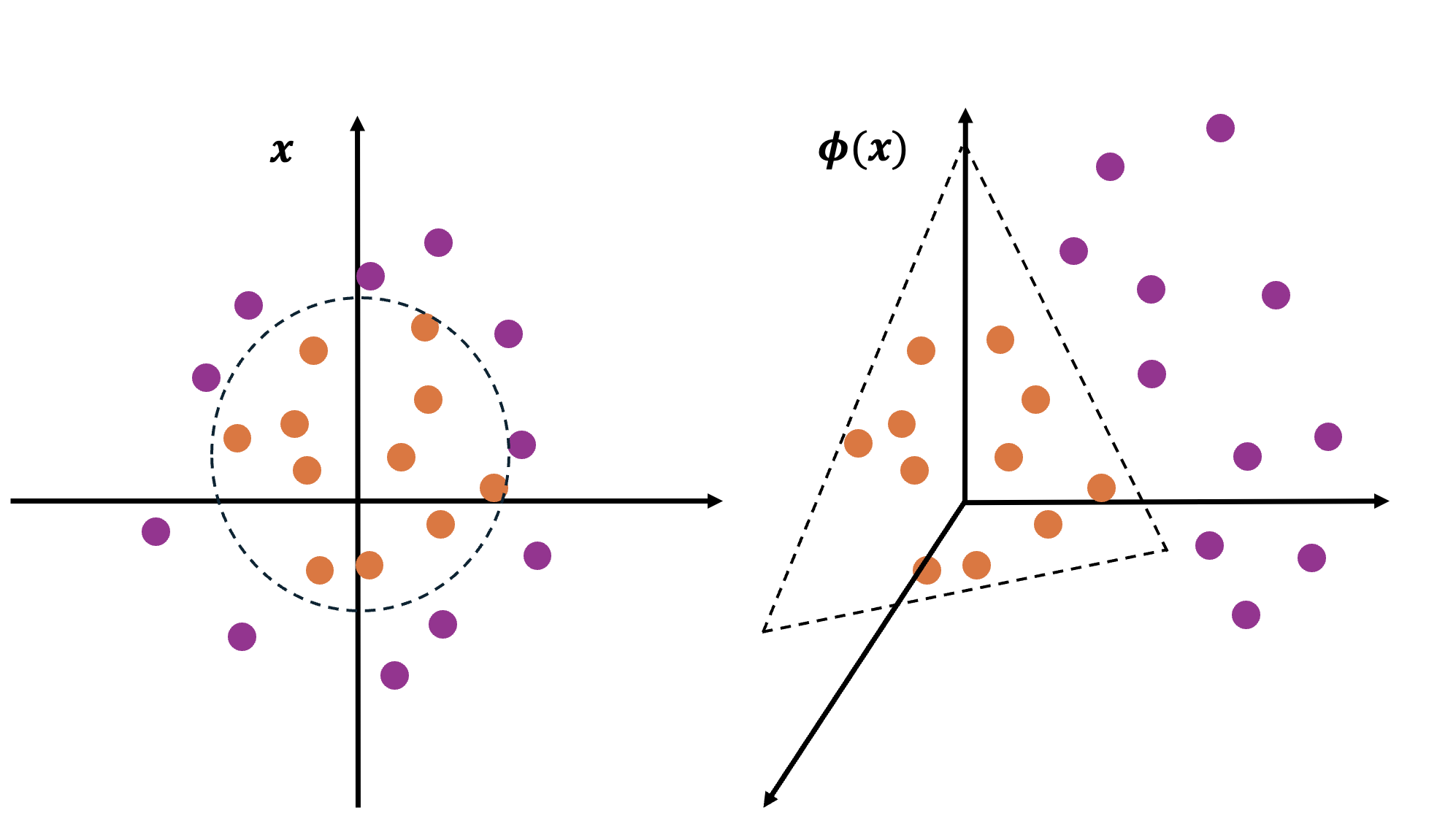}
 \caption{An example of overlapping distribution in the original feature space $\textbf{x}$ and the same sample in higher dimensional feature space $\Phi(\textbf{x})$, where the distribution is linearly separable.}
 \label{fig:fig3}
 \end{figure}

Many `real-world' classification problems are not linearly separable. In this section, we explore how SVMs address non-linearly separable data using the `kernel trick'. Figure~\ref{fig:fig3} presents the same dataset in two different scenarios: on the left, two classes are shown in a lower-dimensional space, denoted as $\textbf{x}$, where the datasets are not linearly separable. On the right, the same distribution is mapped into a higher-dimensional feature space where it becomes linearly separable. This transformation demonstrates that a distribution which is not linearly separable in a low-dimensional space can become separable in a higher-dimensional space with the appropriate transformation, as shown in~\cite{boser1992training,cortes1995support,hearst1998support}.

In both hard and soft margin scenarios, the optimization process in an SVM depends on the scalar product of feature vectors, as illustrated in Equation~\ref{eq:eq3}. Let $\Phi$ represent the transformation function that maps the low-dimensional original feature space to a higher-dimensional space, described as $\Phi: \textbf{x} \rightarrow \Phi(\textbf{x})$. Following this transformation, the inner product component in the Lagrangian maximization equation is modified accordingly, as shown in Equation~\ref{eq:eq6}. However, a significant challenge arises when this transformation leads to a very high-dimensional space. In such cases, computing the inner product becomes computationally intensive, requiring the calculation of each transformed vector $\Phi(\textbf{x}_{i})$ and then determining their inner product in the expanded space. The `kernel trick' offers an powerful solution to this problem by simplifying the computation of these inner products, without explicitly performing the high-dimensional transformation.  

\begin{equation}
\label{eq:eq6}
    \textbf{x}_{i}.\textbf{x}_{j} = K(\textbf{x}_{i},\textbf{x}_{j}) =\Phi(\textbf{x}_{i}).\Phi(\textbf{x}_{j})
\end{equation}

If there exits a kernel function $K(\textbf{x}_{i},\textbf{x}_{j})$ such that Equation~\ref{eq:eq6} is satisfied, then the explicit calculation of $\Phi$ is not required. The classification function for an unknown data $\textbf{u}$ would then take the form indicated by Equation~\ref{eq:eq7}, analogous to the scenario where the data is linearly separable. Identifying a suitable function that meets the criteria of a kernel can be challenging. However, if a function is positive semi-definite and satisfies the conditions set forth by Mercer's theorem, it is guaranteed to be a valid kernel function~\cite{mercer1909xvi}.

\begin{equation}
    \label{eq:eq7}
  f(x) = \sum_{s\in SV} \alpha_{s}y_{s}K(\textbf{x}_{i}.\textbf{u}) + b 
 \end{equation}

Some of the commonly used kernels include: 
\begin{itemize}
    \item Polynomial Kernel (PK): $K(\textbf{x}_{i},\textbf{x}_{j})$ = $(1+\textbf{x}_{i}.\textbf{x}_{j})^{p}$, p is the order of the polynomial 
    \item Gaussian Kernel (GK): $K(\textbf{x}_{i},\textbf{x}_{j})$ = $exp(-\frac{||\textbf{x}_{i}-\textbf{x}_{j}||^{2}}{2\sigma^{2}})$, with $\sigma$ >0 being a user-choosen parameter
    \item Sigmoid Kernel (SK): $K(\textbf{x}_{i},\textbf{x}_{j})$ = $\tanh(k\textbf{x}_{i}.\textbf{x}_{j} - \delta)$, k and $\delta$ are user-choosen kernel parameters
\end{itemize}
The choice of kernel in classification problems is application-dependent. Typically, the Gaussian kernel outperforms the polynomial kernel in terms of accuracy and convergence time \cite{BW10}. The Gaussian kernel is notable for its interpolation ability and effectiveness in capturing local properties, but it has computational and scalability issues, particularly with large datasets. In contrast, the sigmoid kernel is better suited to identify global characteristics but exhibits relatively weak interpolation ability \cite{yan2023reconfigurable}. To leverage the advantages of both types of kernels, we employ a custom kernel, mixed-kernel, as represented by Equation~\ref{eq:eq8} with a linear superposition of Gaussian and Sigmoid kernel. The parameter (\textit{r}) represents the mixing ratio between the components of the Sigmoid and Gaussian kernel. The performance of this mixed kernel approach for charge cluster classification in pixel detectors of trackers in HEP experiments will be discussed in Section~\ref{sec:mkp}.

\begin{equation}
\label{eq:eq8}
\begin{split}
    &\text{mixed-kernel} = (1-r)\times \text{SK} + r\times \text{GK}, \quad 0 \leq r \leq 1 \\
    \text{mixed-kernel} &= (1-r)\times \left[\tanh\left(k\mathbf{x}_{i} \cdot \mathbf{x}_{j} - \delta\right)\right] + r\times \left[\exp\left(-\gamma \|\mathbf{x}_{i} - \mathbf{x}_{j}\|^{2}\right)\right], \quad \gamma = \frac{1}{2\sigma^{2}}
\end{split}
\end{equation}

\section{Mixed-Kernel SVM Model in Tracking}
\label{sec:mkp}

Tracking detectors play a crucial role in capturing primary interactions during beam collisions and generate a substantial volume of data. A significant portion of these data come from tracks produced by low-transverse momentum particles, which are often irrelevant to high-energy collisions of interest. Therefore, efficiently identifying and rejecting these low$p_T$ tracks in real-time is essential to effectively manage data volumes. In this section, we will investigate the performance of SVM using a mixed-kernel approach to filter out these low-$p_T$ charge clusters.

\subsection{Pixel Detector Data}
\label{subsec:spd}
\begin{figure}[htbp]
 \centering
  \includegraphics[width= 0.46 \linewidth]{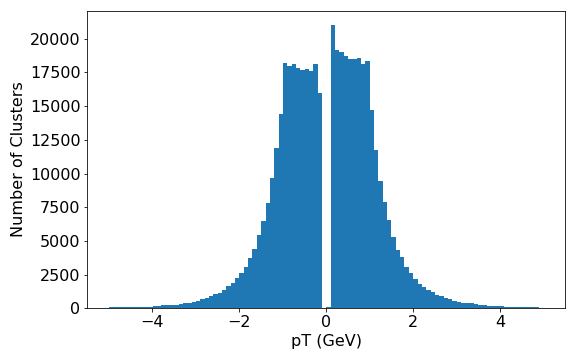}
  \includegraphics[width= 0.52 \linewidth]{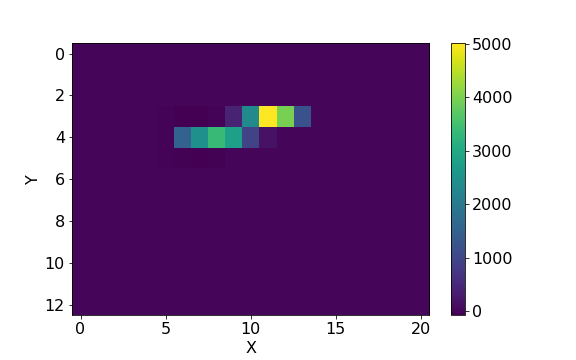}
 
 \caption{(Left) $p_T$ distribution of positive and negative pions. (Right) Hit-map displaying the charge deposition pattern across the 21×13 array of pixel sensors.}
 \label{fig:fig4}
 \end{figure}

In our study, we used the simulated data of smart pixel sensors, which is the charge deposited in the silicon sensors due to interactions of pions~\cite{zenodo_record}. This dataset provides detailed information on the charge deposited in each pixel for every time slice. These pixels form part of a 21×13 array, representing a selected region of interest from the larger 16 mm × 16 mm flat silicon sensor. The $p_T$ distribution of the charged pions and hit map of the charge deposited by a single pion in pixel array is shown in Figure~\ref{fig:fig4}. The accumulated charge in pixel rows, projected onto the y-axis, forms what is termed as the \textit{y-profile}. The SVM model inputs 14 features for each track, with 13 features derived from the \textit{y-profile} and one from the $y_0$, which is the distance in $y$-axis between impact point and center of the flat module. The labeling of the charge clusters was done based on the $p_T$ values of the incident pions. The charge clusters produced by the pions with |$p_T$| > 0.2 GeV are labeled as high-$p_T$ samples, and otherwise as low-$p_T$ samples. 

\subsection{Model Architecture and Optimization}
\label{subsec:svma}

\begin{figure}[htbp]
 \centering
 \includegraphics[width= \linewidth]{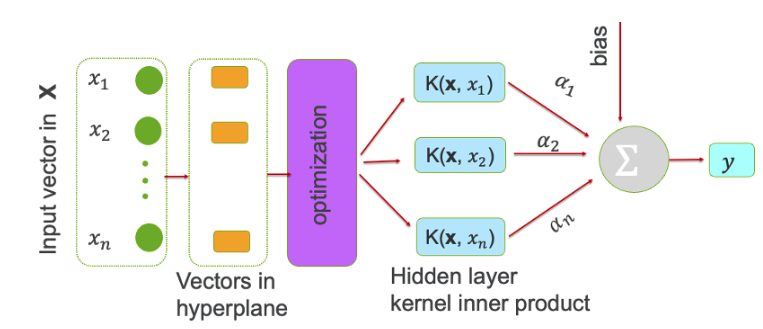}
 \caption{The schematic diagram of the SVM model architecture with mixed-kernel approach. The output is the label predicted by the model for each charge track in the pixel detector.}
 \label{fig:fig5}
 \end{figure}

The schematic diagram in Figure~\ref{fig:fig5} shows the architecture of the SVM model according to ref.~\cite{deo2016wavelet}. The model accepts the input vector \textbf{X}, consisting of 14 features ($x_1, x_2, \ldots, x_{14}$) for each charge cluster, which represent the total charge per sensor. These features undergo a transformation mapping onto vectors in the hyperplane to find the optimum separating hyperplane. However, employing the kernel trick allows for the computation of the inner products between support vectors without necessitating the transformation of the entire input feature set into a high-dimensional space. 

The SVM model with the mixed-kernel approach requires optimization of four parameters ({$r, k, \delta, \gamma$}). Furthermore, there is an additional parameter, the regularization factor (\textit{\textbf{C}}), which plays a critical role in the model's complexity control~\cite{scholkopf2002learning}. The optimization of these five hyperparameters was performed using an autotuning software package, \textit{ytopt}~\cite{wu2023autotuning, wu2024autotuning}. \textit{ytopt} uses Bayesian optimization to find the best input parameter configuration for a given kernel. An accuracy for predicting high-$p_T$ and low-$p_T$ samples serves as the metric guiding the optimization process.  The kernel function is associated with a Lagrange multiplier $\alpha_{i}$, which determines the weight of the corresponding support vector in the decision function. The cumulative decision function is then computed by summing the weighted kernel functions and adding a bias term. The final output \textit{y} is the predicted class label of the input vector \textbf{X}, which is a 14 feature charge cluster.


\subsection{Performance Evaluation}
\label{subsec:per}
In the field of high energy physics, signals that carry vital information, such as the decay of Higgs bosons, are often exceedingly rare compared to background events. Accordingly, our evaluation of the SVM model's performance focuses on metrics that reflect its ability to accurately classify high-$p_T$ signals (signals) thereby ensuring high signal acceptance with minimal false negatives. Concurrently, the model must effectively reject less important low-$p_T$ samples (background) to maintain data volumes within manageable limits. Therefore, in addition to regular classification metrics such as accuracy, precision, recall, and F1score, we considered the primary emphasis on the following two metrics to study the performance of the SVM model with the mixed-kernel approach.
\begin{itemize}
    \item Signal Efficiency: This metric represents the proportion of charge clusters with |$p_T$| > 2 GeV that are correctly classified by the model as high-$p_T$ samples
    
    \item Background Rejection: This is the fraction of charge clusters with |$p_T$| < 2 GeV that are accurately identified by the model as low-$p_T$ samples
\end{itemize}

\begin{figure}[htbp]
 \centering
 \includegraphics[width= \linewidth]{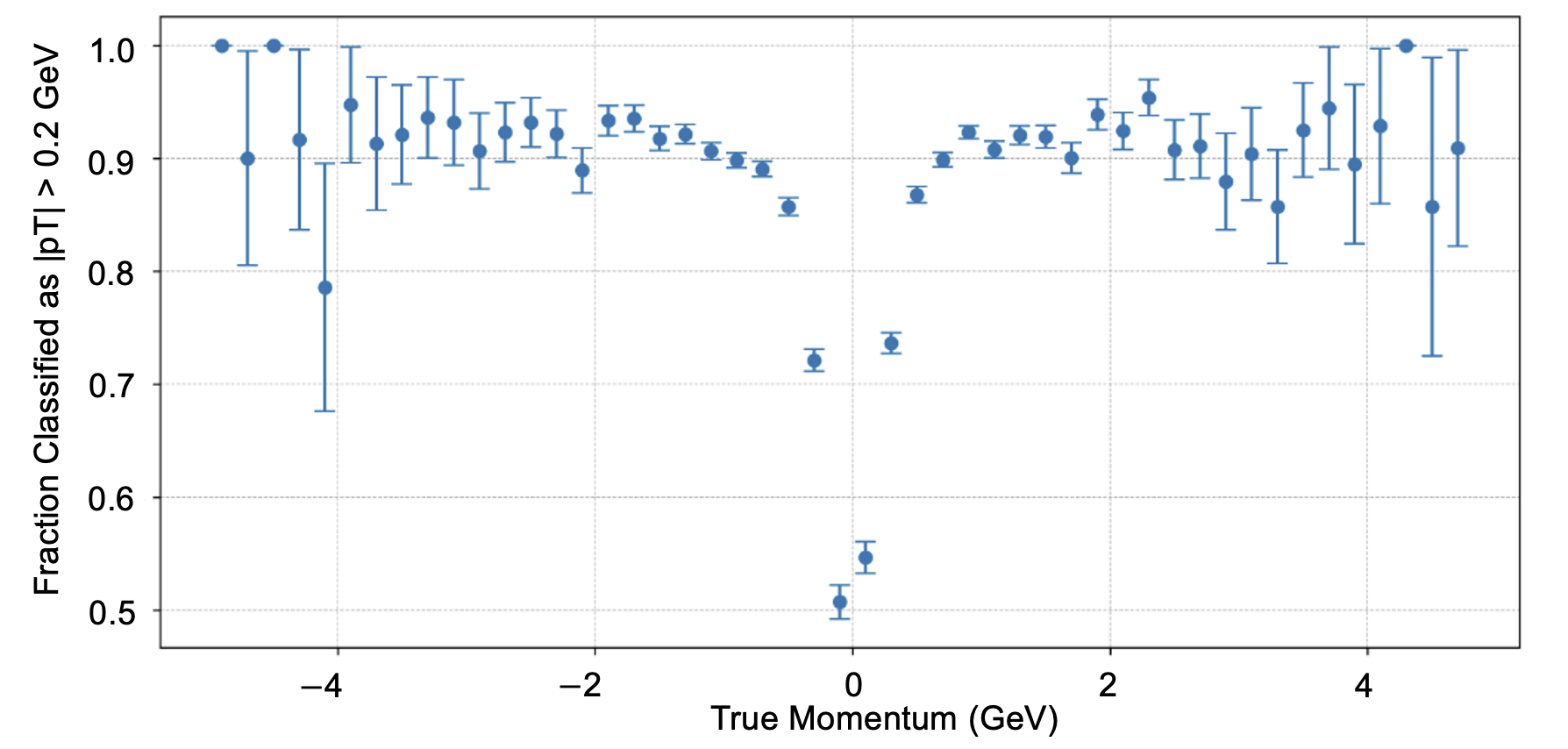}
 \caption{Fraction of charge clusters classified as high-$p_T$ samples as a function of their true transverse momentum (true-$p_T$) values. The distinction between positive and negative $p_T$ values corresponds to the charge of the pions (positive or negative). On the X-axis, the true-$p_T$ is binned with a width of 0.2 GeV. The error bars represent the statistical uncertainty based on the number of charge clusters within each 0.2 GeV bin.}
 \label{fig:fig6}
 \end{figure}

\begin{table}[ht]
\centering
\begin{tabular}{|c|c|c|}
\hline
Datasets & Fraction of Datasets & Background Rejection \\
\hline
Track data & 40\% & 15.9\% \\
Untrack data & 55\% & 50.6\% \\
Single pixel hit data & 5\% & 100\% \\
\hline
\end{tabular}
\caption{Summary of dataset characteristics and background rejection rates.}
\label{tab:table 1}
\end{table}

Figure~\ref{fig:fig6} shows the fraction of tracks classified as high-$p_T$ vs the true momentum of the tracks. The implementation of the mixed-kernel SVM model has achieved a signal efficiency of 91.7\% with a background rejection rate of 15.9\% in the track data. Of the remaining data, 55\% is untracked, primarily from low-$p_T$ tracks, and other detector effects~\cite{yoo2024smart}. The model can reject 50.6\% of the untracked data. Table~\ref{tab:table 1} shows the data description, fraction of each type of data in real experiment, and corresponding background rejection that can be achieved. Additionally, the remaining 5\% of the data, originating from the single pixel hits, can be directly rejected. Consequently, the overall background rejection of 39.2\% could be achieved while maintaining signal efficiency of 91.7\%. These performance metrics highlight the model’s potential as an instrumental tool for data reduction in tracking detectors, while maintaining a high rate of signal capture. 

\section{Mixed-Kernel SVM Device and Co-Design Adaptation}
\label{sec:svmd}
Our study demonstrates that the SVM model with a mixed-kernel approach is highly effective in classifying charge clusters from silicon tracking detectors in HEP experiments. Integrating such a model into hardware could facilitate real-time filtering of low-$p_T$
clusters in these experiments, increasing the relevancy of data sent to downstream analysis systems. A notable device, the mixed-kernel heterojunction (MKH) transistor, was developed for a similar task, the real-time detection of arrhythmias from electrocardiogram signals (described in \cite{yan2023reconfigurable}). The MKH transistor is constructed from the low-dimensional materials MoS$_2$ and carbon nanotubes, and is distinguished by its reconfigurable nature which allows it to produce a range of electronic characteristics that resemble the mixed Gaussian-Sigmoidal kernel. Additionally, its utilization of low-dimensional materials gives it minimal radiation cross-section that offers significant advantages in terms of area and radiation constraints in experimental settings. In this section, we will discuss the MKH transistor's power consumption, potential adaptations, and limitations as part of a hardware implementation of a mixed-kernel SVM.

\subsection{Power Consumption}
\label{subsec:pc}
\begin{figure}[htbp]
    \centering
    \includegraphics[width= \linewidth]{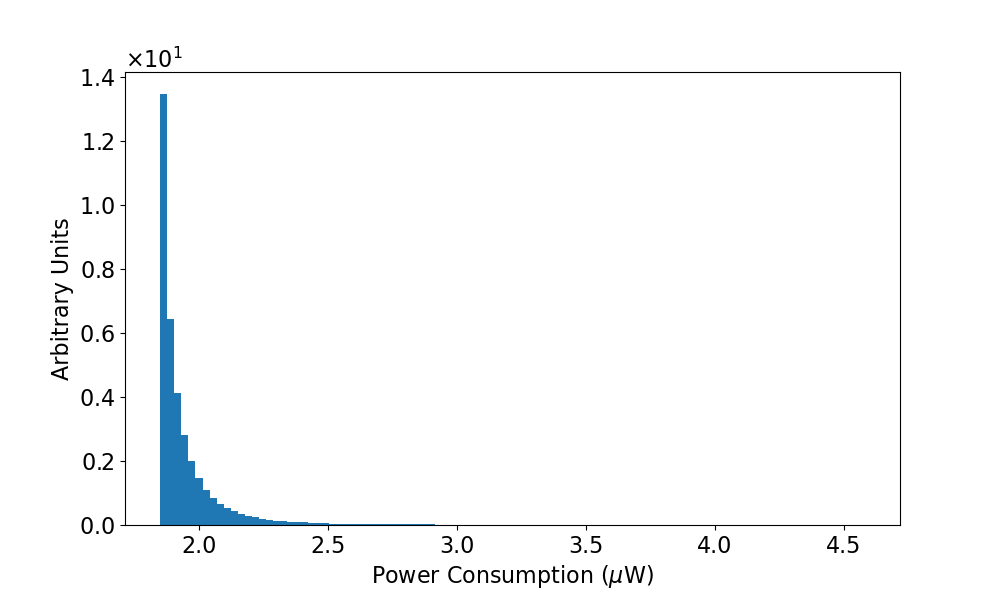}
    \caption{Power consumption of the reconfigurable mixed-kernel SVM device~\cite{yan2023reconfigurable} for classifying low-$p_T$ and high-$p_T$ charge clusters from the pixel detectors }
    \label{fig:fig8}
\end{figure}

The MKH transistor enables a complete set of Gaussian, Sigmoid and mixed-kernel with only a single device, which requires reduced circuitry requirements, making it significantly more power-efficient by approximately two orders of magnitude compared to traditional analog CMOS circuits designed for similar kernel generation \cite{yan2023reconfigurable}. To predict the amount of power consumed by the device for classifying the tracks, the dynamic range of input signals are scaled to the corresponding region of operation for the MKH transistor. Figure~\ref{fig:fig8} shows the distribution of power consumption for an MKH transistor carrying out the kernel transformation in an SVM classification task. The estimated mean power consumption for the kernel is low, at about 1.97$\mu$W. 

\begin{figure}[htbp]
    \centering
    \includegraphics[width= \linewidth]{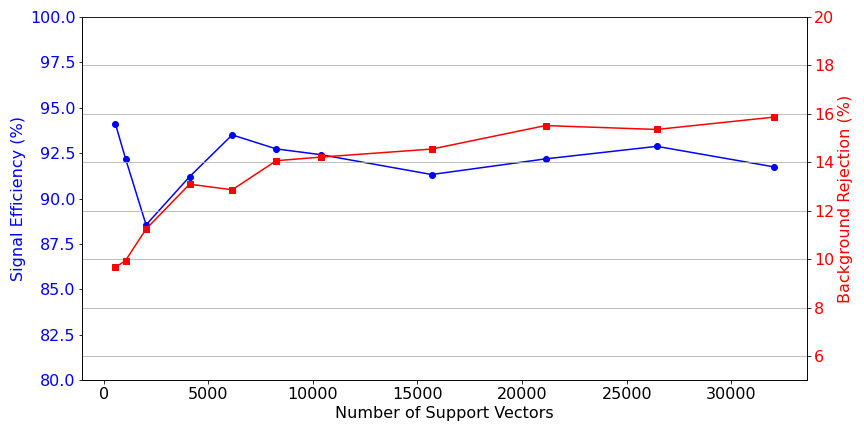}
    \caption{Performance metrics versus the number of support vectors. The left Y-axis represents signal efficiency, and the right Y-axis indicates the background rejection. The results correspond to the tracked data.}
    \label{fig:fig10}
\end{figure}

While the power consumed by the kernel operation is low, this operation must be preceded by the dot-product of the input vector to be classified with the support vectors which define the decision boundary of the kernelized SVM (\ref{eq:eq7}). To carry out this preceding operation in the analog domain, techniques such as crossbar arrays of resistive memories have been demonstrated as a feasible approach to store and carry out low-energy dot products with support vectors \cite{Lee_Jeon_Park_Yoo_Kim_Ha_Hwang_2019, Xia_Gu_Li_Tang_Yin_Huangfu_Yu_Cao_Wang_Yang_2016}. The performance of the mixed-kernel SVM classifier varies with the number of support vectors used to construct its decision boundary (Fig. \ref{fig:fig10}), creating a trade-off between the amount of power, area, and performance of the system based on the number of support vectors it employs. Currently, we have demonstrated an opportunity for co-design in this system based on tuning a mixed kernel to a given application which is implemented in a single device; further investigation of the trade-off between support vectors, circuit area, and performance is left for future work. The mixed-kernel SVM device, operating in the analog domain, introduces a novel approach to on-detector data processing in HEP experiments. 

\section{Summary}
\label{sec:discuss}
Given the unprecedented data rates expected in future HEP experiments due to increased luminosity and detector size, an on-detector system for data processing and the rejection of unwanted data is essential to keep the data volume manageable. This work presents the performance and feasibility of support vector machine models with a mixed-kernel approach for data processing in high-energy physics experiments. The mixed-kernel, which combines Sigmoid and Gaussian kernels, leverages the advantage of both kernels and outperforms in the performance metrics. We have demonstrated that the mixed-kernel SVM exhibits excellent performance in classifying charge clusters from silicon pixel sensors, which are a crucial part of the tracking detector system in HEP experiments. Our findings indicate that, overall, approximately 39\% of data could be rejected in real-time while maintaining a signal efficiency of about 92\% with a mixed-kernel SVM approach. 

The 2D heterojunction tunable device, the MKH transistor, offers a promising solution for on-detector data processing within the tracker system of HEP experiments. We demonstrate that the device is particularly energy-efficient for the kernel generation of mixed-kernel SVM classifiers. Further work is needed to evaluate the integration of this device with technologies capable of carrying out dot-products for the construction of cohesive, low-power, analog-domain SVM classifiers capable of computing with mixed kernels.


\acknowledgments
We would like to thank Jinlong Zhang, Farah Fahim, Nhan Tran, Jennet Dickinson, and Alexander Paramov for many fruitful discussions which guided this work. 

This work was supported by DOE ASCR and BES
Microelectronics Threadwork. This material is based upon work
supported by the U.S. Department of Energy, Office of Science,
under contract number DE-AC02-06CH11357. The 2D heterojunction work was also partially supported by the National Science Foundation Materials Research Science and Engineering Center at Northwestern University under Award Number DMR-2308691


\bibliographystyle{JHEP}
\bibliography{biblio.bib}

\providecommand{\href}[2]{#2}\begingroup\raggedright\begin{thebibliography}{10}

\bibitem{alam2008atlas}
S.~Alam, B.~Athar, S.~Timm, F.~Wappler, L.~Zhichao, H.~Ahmed et~al., \emph{The atlas experiment at the cern large hadron collider}, \href{https://doi.org/10.1088/1748-0221/3/08/S08003}{\emph{Journal of instrumentation} {\bfseries 3} (2008) S08003}.

\bibitem{collaboration2008cms}
C.~Collaboration, S.~Chatrchyan, G.~Hmayakyan, V.~Khachatryan, A.~Sirunyan, W.~Adam et~al., \emph{The cms experiment at the cern lhc}, \href{https://doi.org/10.1088/1748-0221/3/08/S08004}{\emph{JInst} {\bfseries 3} (2008) S08004}.

\bibitem{DellaNegra_Jenni_Virdee_2012}
M.~Della~Negra, P.~Jenni and T.S.~Virdee, \emph{Journey in the search for the higgs boson: The atlas and cms experiments at the large hadron collider}, \href{https://doi.org/10.1126/science.1230827}{\emph{Science} {\bfseries 338} (2012) 1560–1568}.

\bibitem{Das_2022}
P.~Das, \emph{An overview of the trigger system at the cms experiment}, \href{https://doi.org/10.1088/1402-4896/ac6302}{\emph{Physica Scripta} {\bfseries 97} (2022) 054008}.

\bibitem{fcc2019fcc}
F.~collaboration et~al., \emph{Fcc-ee: the lepton collider: future circular collider conceptual design report volume 2}, \href{https://doi.org/10.1140/epjst/e2019-900045-4}{\emph{European Physical Journal Special Topics} {\bfseries 228} (2019) 261}.

\bibitem{fcc2019hh}
F.~collaboration et~al., \emph{Fcc-hh: the hadron collider: future circular collider conceptual design report volume 3}, \href{https://doi.org/https://doi.org/10.1140/epjst/e2019-900087-0}{\emph{European Physical Journal Special Topics} {\bfseries 228} (2019) 755}.

\bibitem{vazquez2016atlas}
W.P.~Vazquez, A.~Collaboration et~al., \emph{The atlas data acquisition system: from run 1 to run 2}, \href{https://doi.org/10.1016/j.nuclphysbps.2015.09.146}{\emph{Nuclear and particle physics proceedings} {\bfseries 273} (2016) 939}.

\bibitem{tdaq}
{ATLAS Collaboration}, ``Daq approved plots.'' \url{https://twiki.cern.ch/twiki/bin/view/AtlasPublic/ApprovedPlotsDAQ}, 2017.

\bibitem{yoo2024smart}
J.~Yoo, J.~Dickinson, M.~Swartz, G.~Di~Guglielmo, A.~Bean, D.~Berry et~al., \emph{Smart pixel sensors: towards on-sensor filtering of pixel clusters with deep learning}, \href{https://doi.org/10.1088/2632-2153/ad6a00}{\emph{Machine Learning: Science and Technology} (2024) }.

\bibitem{bartz2020fpga}
E.~Bartz, G.~Boudoul, R.~Bucci, J.~Chaves, E.~Clement, D.~Cranshaw et~al., \emph{Fpga-based tracking for the cms level-1 trigger using the tracklet algorithm}, \href{https://doi.org/10.1088/1748-0221/15/06/P06024}{\emph{Journal of Instrumentation} {\bfseries 15} (2020) P06024}.

\bibitem{boser1992training}
B.E.~Boser, I.M.~Guyon and V.N.~Vapnik, \emph{A training algorithm for optimal margin classifiers},  in \emph{Proceedings of the fifth annual workshop on Computational learning theory}, pp.~144--152, 1992, \href{https://doi.org/https://doi.org/10.1145/130385.13040}{DOI}.

\bibitem{cortes1995support}
C.~Cortes and V.~Vapnik, \emph{Support-vector networks}, \href{https://doi.org/10.1007/BF00994018}{\emph{Machine learning} {\bfseries 20} (1995) 273}.

\bibitem{awad2015support}
M.~Awad, R.~Khanna, M.~Awad and R.~Khanna, \emph{Support vector regression}, \href{https://doi.org/10.1007/978-1-4302-5990-9}{\emph{Efficient learning machines: Theories, concepts, and applications for engineers and system designers} (2015) 67}.

\bibitem{gunn1998support}
S.R.~Gunn et~al., \emph{Support vector machines for classification and regression}, {\emph{ISIS technical report} {\bfseries 14} (1998) 5}.

\bibitem{zhang2008text}
W.~Zhang, T.~Yoshida and X.~Tang, \emph{Text classification based on multi-word with support vector machine}, \href{https://doi.org/10.1016/j.knosys.2008.03.044}{\emph{Knowledge-Based Systems} {\bfseries 21} (2008) 879}.

\bibitem{yang2004biological}
Z.R.~Yang, \emph{Biological applications of support vector machines}, \href{https://doi.org/10.1093/bib/5.4.328}{\emph{Briefings in bioinformatics} {\bfseries 5} (2004) 328}.

\bibitem{lin2011large}
Y.~Lin, F.~Lv, S.~Zhu, M.~Yang, T.~Cour, K.~Yu et~al., \emph{Large-scale image classification: Fast feature extraction and svm training},  in \emph{CVPR 2011}, pp.~1689--1696, IEEE, 2011, \href{https://doi.org/10.1109/CVPR.2011.5995477}{DOI}.

\bibitem{shin2005application}
K.-S.~Shin, T.S.~Lee and H.-j.~Kim, \emph{An application of support vector machines in bankruptcy prediction model}, \href{https://doi.org/10.1016/j.eswa.2004.08.009}{\emph{Expert systems with applications} {\bfseries 28} (2005) 127}.

\bibitem{chen2011support}
H.-L.~Chen, B.~Yang, J.~Liu and D.-Y.~Liu, \emph{A support vector machine classifier with rough set-based feature selection for breast cancer diagnosis}, \href{https://doi.org/10.1016/j.eswa.2011.01.120}{\emph{Expert systems with applications} {\bfseries 38} (2011) 9014}.

\bibitem{ahmad2018performance}
I.~Ahmad, M.~Basheri, M.J.~Iqbal and A.~Rahim, \emph{Performance comparison of support vector machine, random forest, and extreme learning machine for intrusion detection}, \href{https://doi.org/10.1109/ACCESS.2018.2841987}{\emph{IEEE access} {\bfseries 6} (2018) 33789}.

\bibitem{vannerem1999classifying}
P.~Vannerem, K.-R.~M{\"u}ller, B.~Sch{\"o}lkopf, A.~Smola and S.~Soldner-Rembold, \emph{Classifying lep data with support vector algorithms}, \href{https://doi.org/10.48550/arXiv.hep-ex/9905027}{\emph{arXiv preprint hep-ex/9905027} (1999) }.

\bibitem{vaiciulis2003support}
A.~Vaiciulis, \emph{Support vector machines in analysis of top quark production}, \href{https://doi.org/10.1016/S0168-9002(03)00479-0}{\emph{Nuclear Instruments and Methods in Physics Research Section A: Accelerators, Spectrometers, Detectors and Associated Equipment} {\bfseries 502} (2003) 492}.

\bibitem{aaltonen2012search}
T.~Aaltonen, B.{\'A}.~Gonz{\'a}lez, S.~Amerio, D.~Amidei, A.~Anastassov, A.~Annovi et~al., \emph{Search for the standard model higgs boson produced in association with a w$\pm$boson with 7.5 fb- 1 integrated luminosity at cdf}, \href{https://doi.org/10.1103/PhysRevD.86.032011}{\emph{Physical Review D} {\bfseries 86} (2012) 032011}.

\bibitem{sahin2016performance}
M.{\"O}.~Sahin, D.~Kr{\"u}cker and I.-A.~Melzer-Pellmann, \emph{Performance and optimization of support vector machines in high-energy physics classification problems}, \href{https://doi.org/10.1016/j.nima.2016.09.017}{\emph{Nuclear Instruments and Methods in Physics Research Section A: Accelerators, Spectrometers, Detectors and Associated Equipment} {\bfseries 838} (2016) 137}.

\bibitem{hearst1998support}
M.A.~Hearst, S.T.~Dumais, E.~Osuna, J.~Platt and B.~Scholkopf, \emph{Support vector machines}, \href{https://doi.org/10.1109/5254.708428}{\emph{IEEE Intelligent Systems and their applications} {\bfseries 13} (1998) 18}.

\bibitem{scholkopf2002learning}
B.~Sch{\"o}lkopf and A.J.~Smola, \emph{Learning with kernels: support vector machines, regularization, optimization, and beyond}, MIT press (2002), \href{https://doi.org/10.7551/mitpress/4175.001.0001}{10.7551/mitpress/4175.001.0001}.

\bibitem{shawe2011review}
J.~Shawe-Taylor and S.~Sun, \emph{A review of optimization methodologies in support vector machines}, \href{https://doi.org/10.1016/j.neucom.2011.06.026}{\emph{Neurocomputing} {\bfseries 74} (2011) 3609}.

\bibitem{suykens2001nonlinear}
J.A.~Suykens, \emph{Nonlinear modelling and support vector machines},  in \emph{IMTC 2001. proceedings of the 18th IEEE instrumentation and measurement technology conference. Rediscovering measurement in the age of informatics (Cat. No. 01CH 37188)}, vol.~1, pp.~287--294, IEEE, 2001, \href{https://doi.org/10.1109/IMTC.2001.928828}{DOI}.

\bibitem{mercer1909xvi}
J.~Mercer, \emph{Xvi. functions of positive and negative type, and their connection the theory of integral equations}, \href{https://doi.org/10.1098/rsta.1909.0016}{\emph{Philosophical transactions of the royal society of London. Series A, containing papers of a mathematical or physical character} {\bfseries 209} (1909) 415}.

\bibitem{BW10}
A.~Ben-Hur and J.~Weston, \emph{A user's guide to support vector machines},  in \emph{Data Mining Techniques for the Life Sciences}, O.~Carugo and F.~Eisenhaber, eds., (Totowa, NJ), pp.~223--239, Humana Press (2010), \href{https://doi.org/10.1007/978-1-60327-241-4_13}{DOI}.

\bibitem{yan2023reconfigurable}
X.~Yan, J.H.~Qian, J.~Ma, A.~Zhang, S.E.~Liu, M.P.~Bland et~al., \emph{Reconfigurable mixed-kernel heterojunction transistors for personalized support vector machine classification}, \href{https://doi.org/10.1038/s41928-023-01042-7}{\emph{Nature Electronics} {\bfseries 6} (2023) 862}.

\bibitem{zenodo_record}
D.~Berry et~al., ``Smart pixel dataset.''
\newblock 10.5281/zenodo.7331127.

\bibitem{deo2016wavelet}
R.C.~Deo, X.~Wen and F.~Qi, \emph{A wavelet-coupled support vector machine model for forecasting global incident solar radiation using limited meteorological dataset}, \href{https://doi.org/10.1016/j.apenergy.2016.01.130}{\emph{Applied Energy} {\bfseries 168} (2016) 568}.

\bibitem{wu2023autotuning}
X.~Wu, P.~Paramasivam and V.~Taylor, \emph{Autotuning apache tvm-based scientific applications using bayesian optimization},  in \emph{Proceedings of the SC'23 Workshops of The International Conference on High Performance Computing, Network, Storage, and Analysis}, pp.~29--35, 2023, \href{https://doi.org/https://doi.org/10.1145/3624062.3626079}{DOI}.

\bibitem{wu2024autotuning}
X.~Wu, T.~Oli, V.~Taylor, M.C.~Hersam, V.K.~Sangwan et~al., \emph{An autotuning-based optimization framework for mixed-kernel svm classifications in smart pixel datasets and heterojunction transistors}, \href{https://doi.org/10.48550/arXiv.2406.18445}{\emph{arXiv preprint arXiv:2406.18445} (2024) }.

\bibitem{Lee_Jeon_Park_Yoo_Kim_Ha_Hwang_2019}
Y.K.~Lee, J.W.~Jeon, E.-S.~Park, C.~Yoo, W.~Kim, M.~Ha et~al., \emph{Matrix mapping on crossbar memory arrays with resistive interconnects and its use in in-memory compression of biosignals}, \href{https://doi.org/10.3390/mi10050306}{\emph{Micromachines} {\bfseries 10} (2019) 306}.

\bibitem{Xia_Gu_Li_Tang_Yin_Huangfu_Yu_Cao_Wang_Yang_2016}
L.~Xia, P.~Gu, B.~Li, T.~Tang, X.~Yin, W.~Huangfu et~al., \emph{Technological exploration of rram crossbar array for matrix-vector multiplication}, \href{https://doi.org/10.1007/s11390-016-1608-8}{\emph{Journal of Computer Science and Technology} {\bfseries 31} (2016) 3–19}.

\end{thebibliography}\endgroup

\end{document}